\documentclass[12pt]{iopart} 
\usepackage{epsfig}                   

\begin{document} 
\title[Gluino Contribution to Radiative $B$ Decays]{Gluino
 Contribution to Radiative $B$ Decays: \\
 New Operators, Organization of QCD
 Corrections and Leading Order Results~\footnote{Presented by D.~Wyler 
 at ``Heavy Flavour 2000'' (Durham); based on Ref.~\cite{borz}}}
\author{Daniel Wyler$^a$ and Francesca Borzumati$^b$}
\address{
$^a$ Institut f\"ur Theoretische Physik, Universit\"at Z\"urich,
 CH--8057 Zurich, Switzerland \\
%
$^b$ SISSA, Via Beirut 4, I--34014, Italy}

\begin{abstract}
 The gluino-induced contribution to the decay $b \to s \gamma$ is
 investigated in supersymmetric frameworks with generic sources of
 flavour violation. It is emphasized that the operator basis of
 the standard model effective Hamiltonian is enlarged and that a
 suitable redefinition allows to realize the usual scheme of leading
 and next-to-leading logarithmic contributions when QCD 
 corrections are included.  The effects of the leading order
 QCD corrections on the inclusive branching ratio for $b \to s \gamma$
 are shown. Constraints on supersymmetric sources of flavour violation
 are derived.
\end{abstract}
\section{Introduction}
Like the usual weak interactions, also new physics interactions give
rise to an `effective' Hamiltonian operator at low energies (well below
$M_W$).  It only contains the `active' particles, that is quarks,
leptons, photons and gluons. The details of the heavy (new) particles
and forces on them acting disappear and their properties manifest
themselves in the form and the strength of the various operators in the
effective Hamiltonian.

In the standard model (SM), the only operator at tree level comes from
$W$-exchange:
\begin{equation}
\label{forf}
{\cal H}_{eff} \sim 
G_F (\bar q_L \gamma_{\mu}q'_L)(\bar q''_L \gamma^{\mu}q'''_L)
\end{equation}
where the $q$ are the various quark flavours and factors like the CKM
matrix elements have been dropped. The form and strength of the
$W$ couplings are embodied in $G_F$ and the left-handedness of the
quarks.  Because of the various gluonic corrections, the number of
active operators is much larger than just the above four-Fermi
term. Dimensional arguments and renormalization group techniques allow
to classify them systematically.

If there are new forces, they will yield not only different couplings
strengths, but also different forms for the interactions. This implies
that there are in general more operators than the one in
eq.~(\ref{forf}) and also more than those obtained in the SM after QCD
corrections have been included.

In supersymmetric models, there are many sources of flavour violations
relevant to the decay $b \to s \gamma$~\cite{BBMR}.  We shall consider
here only those due to a flavour violating gluino-quark-squark
vertex~\cite{DNW}, whose impact on the decay $b \to s \gamma$ was first
analyzed in ref.~\cite{BBM}.  In general, the gluino contribution to
this decay is the largest, unless it is assumed that the only source of
flavour violation is encoded in the superpotential and that the deriving
electroweak-scale flavour-violating parameters in the scalar sector are
small. In this case, the gluino contribution can be neglected, unless
$\tan \beta$ is large~\cite{Borzumati:1994zg}. A complete calculation of
the $b \to s \gamma$ rate, including corrections up to the
next-to-leading order (NLO) )in QCD, exists for a specific class of
models in which the gluino contribution can be
neglected~\cite{Ciuchini:1998xy}. Until recently, no study existed on
the interplay between the gluino contribution to the decay 
$b \to s \gamma$ and QCD corrections, not even at the leading order
(LO), in spite of the potentially important role that the
flavour-violating vertices gluino-quark-squarks may have for model
building (see for example Ref.~\cite{Barenboim:2001ev}).

The gluino contribution to the decay $b \to s \gamma$ exhibits two
special features, when QCD corrections are implemented. First, gluinos
couple to both left- and right-handed quarks. Thus, many more operators
than in the SM, in which only left-handed quarks are coupled to the $W$
boson, are expected. Secondly, the presence of the strong coupling
$\alpha_s$ in the gluino contribution and in the QCD corrections
immediately raises the question of how to order the different powers of
$\alpha_s$ and whether to include them in the gluino-induced operators
or in their Wilson coefficients. These two difficulties necessitate a
dedicated study of the QCD corrections to this contribution.
\section{Ordering the QCD perturbation expansion and the
effective Hamiltonian}
In the SM, rare $B$ decays are induced by loops with $W$ bosons; the
coupling strength is always $G_F$. The corrections are due to gluons and
other light particles and give rise to powers of the large logarithmic
factor $L = log(m^2_b/M^2_W)$. In the decay $b \to s \gamma$ only loops
with gluons contribute, and thus powers of $L$ and $\alpha_s$ are
related as follows~\cite{buras}:\\
\begin{itemize}
 \item[] \quad LO: 
 $\quad \quad G_F \, (\alpha_s L)^N, \quad \quad (N=0,1,...)$  
\item[] \quad NLO:
 $\quad       G_F \, \alpha_s (\alpha_s L)^N$.
\end{itemize}
These terms are summed (over N) using renormalization group
techniques. This results in an effective Hamiltonian
\begin{equation}
 {\cal H}_{eff}^{W} = 
 - \frac{4 G_F}{\sqrt{2}} V_{tb}^{\phantom{\ast}} V_{ts}^\ast
  \sum_i C_i(\mu) {\cal O}_i(\mu) \,, 
\label{weffham}
\end{equation}
where $V_{tb}$ and $V_{ts}$ are elements of the
Cabibbo--Kobayashi--Maskawa (CKM) matrix and $\mu$ is the `scale'. The
Wilson coefficients $C_i$ contain all dependence on the heavy degrees of
freedom, whereas the operators ${\cal O}_{i}$ depend on light fields
only. When all order QCD corrections are included, the Hamiltonian is
independent of the scale $\mu$. At a specific order in QCD, however,
there is a residual dependence on $\mu$. In practice, this scale must be
chosen to be around $m_b$.

The operators relevant to radiative $B$ decays can be divided into two
classes: four-fermion current-current operators and magnetic operators
\begin{equation}
{\cal O}_{M}                 \,= \!
  \displaystyle{\frac{g}{16\pi^2}} \,{\overline m}_b(\mu) \,
 (\bar{s} \sigma^{\mu\nu} P_R b) \, V_{\mu\nu}\,,     
\end{equation}
where $V_{\mu\nu}$ is the field strength of a photon or gluon.  All
these operators are of dimension $six$.

It is by now well known that, in the SM, a consistent calculation for 
$b \to s \gamma$ at LO (or NLO) precision requires three steps:
\begin{itemize}
\item[{\it 1)}] 
a matching calculation of the full SM theory with the effective theory
at the scale $\mu=\mu_W$ to order $\alpha_s^0$ (or $\alpha_s^1$) for
the Wilson coefficients, where $\mu_W$ denotes a scale of order $M_W$ or
$m_t$;
\item[{\it 2)}]  
a renormalization group treatment of the Wilson coefficients using the
anomalous-dimension matrix to order $\alpha_s^1$ (or $\alpha_s^2$);
\item[{\it 3)}]   
a calculation of the operator matrix elements at the scale $\mu = \mu_b$
to order $\alpha_s^0$ (or $\alpha_s^1$), where $\mu_b$ denotes a scale
of order $m_b$.
\end{itemize}

Matters can be different in other $B$ decays or when other contributions
to $b \to s \gamma$ are considered. An example is the decay 
$b \to s \,\ell \, \bar{\ell}$. The first large logarithm
$L=\log(m^2_b/M^2_W)$ arises without the exchange of gluons. This
possibility has no correspondence in the $b\to s \gamma$
case. Consequently, in the case of $b \to s \, \ell \, \bar{\ell}$, the
decay amplitude is ordered according to $ G_F L \, (\alpha_s L)^N$ at
the LO in QCD and $ G_F \alpha_s L (\alpha_s L)^N$ at the NLO.  To
achieve technically the resummation of these terms, it is convenient to
redefine certain operators and their Wilson coefficients as
follows~\cite{M-BM}:
\begin{equation}
\label{reshuffle}
{\cal O}_i^{new} = \frac{16 \pi^2}{g_s^2} {\cal O}_i\,, 
\quad \quad 
C_i^{new} = \frac{g_s^2}{16\pi^2} C_i \quad 
\quad (i=7,...,10). 
\end{equation}
This redefinition allows us to proceed according to the above three
steps when calculating the amplitude of the decay 
$b \to s \, \ell \,\bar{\ell}$~\cite{M-BM}.  In particular, the one-loop
mixing of the operator ${\cal O}_2$ with the operator ${\cal O}_9^{new}$
appears formally at ${\cal O}(\alpha_s)$.

Including gluinos, we can now write the complete effective
Hamiltonian as:
\begin{equation} 
{\cal H}_{eff} = {\cal H}_{eff}^W + {\cal H}_{eff}^{\tilde{g}} \,,
\label{hfull}
\end{equation} 
where $ {\cal H}_{eff}^{W}$ is the SM effective Hamiltonian
in~(\ref{weffham}) and ${\cal H}_{eff}^{\tilde{g}}$ originates after
integrating out squarks and gluinos. Note that `mixed' diagrams, which
contain, besides a $W$ boson, also gluinos and squarks, give rise to
$\alpha_s$ corrections to the Wilson coefficients in ${\cal H}_{eff}^W$
(at the matching scale). Such contributions can be omitted in a LO
calculation, but they have to be taken into account at the NLO level. As
for ${\cal H}_{eff}^{\tilde{g}}$, the aim is to resum the following
terms:
\begin{itemize}
\item[] \quad LO:
 $\quad \quad \alpha_s \, (\alpha_s L)^N, \quad \quad (N=0,1,...)$  
\item[] \quad NLO: 
 $\quad       \alpha_s \, \alpha_s (\alpha_s L)^N$.  
\end{itemize}
\begin{figure}[t]
\vspace*{-0.1truecm}
\begin{center}
\leavevmode
\epsfxsize= 6.0 truecm
\epsfbox[200 560 425 680]{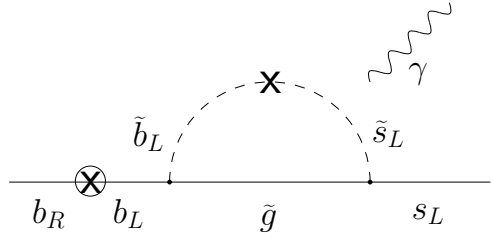}
\end{center}
\caption[f1]{Diagram mediating the $b \to s \gamma$ decay through
 gluino exchange and contributing to the operator 
 ${\cal O}_{7b,\tilde{g}}$. A contribution to the primed operator 
 ${\cal O}_{7b,\tilde{g}}^{\prime}$ is obtained by exchanging
 $L\leftrightarrow R$.} 
\label{glopb}
\end{figure}
While ${\cal H}_{eff}^{\tilde{g}}$ is unambiguous, it is a matter of
convention whether the $\alpha_s$ factors, peculiar of the gluino
exchange, should be put into the definition of operators or into the
Wilson coefficients.  It is convenient (and possible) to distribute
the factors of $\alpha_s$ between operators and Wilson coefficients in
such a way that the first two of the three steps in the program for
the SM calculation also apply to the gluino-induced contribution.
This implies one factor of $\alpha_s^1$ in the definition of the
magnetic and chromomagnetic operators and a factor $\alpha_s^2$ in the
definition of the four-quark operators.  With this convention, the
matching calculation and the evolution down to the low scale $\mu_b$
of the Wilson coefficients are organized exactly in the same way as in
the SM.  The anomalous-dimension matrix, indeed, has the canonical
expansion in $\alpha_s$ and starts with a term proportional to
$\alpha_s^1$.  The last of the three steps in the program of the SM
calculation requires now an obvious modification: the
calculation of the matrix elements has to be performed at order
$\alpha_s$ and $\alpha_s^2$ at the LO and NLO precision.  With this
organization of QCD corrections, the SM Hamiltonian 
${\cal H}_{eff}^{W}$ in eq.~(\ref{weffham}) and the gluino-induced one 
${\cal H}_{eff}^{\tilde{g}}$ undergo separate renormalization, which
facilitates all considerations.

The list of such redefined operators is lengthy and given in 
ref.~\cite{borz}. For illustration only some representative ones
are shown here:

\noindent
$\bullet$ \ 
magnetic operators, with chirality violation coming 
from the $b$-quark mass:
\begin{equation}
{\cal O}_{7b,\tilde{g}}                 \,= \!
   e \,g_s^2(\mu) \,{\overline m}_b(\mu) \,
 (\bar{s} \sigma^{\mu\nu} P_R b) \, F_{\mu\nu}\,,   
\end{equation}
$\bullet$ \ 
magnetic operators in which the chirality-violating parameter is the 
gluino mass $m_{\tilde{g}}$, 
included in the corresponding Wilson coefficients:
\begin{equation}          
{\cal O}_{7\tilde{g},\tilde{g}}         \,= \!
  e \,g_s^2(\mu) \,
 (\bar{s} \sigma^{\mu\nu} P_R b) \, F_{\mu\nu}\,,    
\end{equation}                          
$\bullet$ \ 
magnetic operators, with chirality violation signalled by   
the $c$-quark mass:
\begin{equation}
{\cal O}_{7c,\tilde{g}}                 \,= \!
   e \,g_s^2(\mu) \,{\overline m}_c(\mu) \,
 (\bar{s} \sigma^{\mu\nu} P_R b) \, F_{\mu\nu}\,,   
\end{equation}
$\bullet$ \ 
four-quark operators with vector Lorentz structure:
\begin{equation}
\label{penguinboxop}
{\cal O}_{11,\tilde{g}}^q               \,= \!  g_s^4(\mu) 
(\bar{s} \gamma_\mu  P_L b)\, 
(\bar{q} \gamma^\mu  P_L q) \,,
\end{equation}
$\bullet$ \ 
four-quark operators with scalar and tensor Lorentz structure:
\begin{equation}
\label{boxop}%
{\cal O}_{15,\tilde{g}}^{q}         \,= \!  g_s^4(\mu) 
(\bar{s} P_R b)\,
(\bar{q} P_R q)\,.
\end{equation}
These last four-quark operators are induced by box diagrams only and
through the exchange of two gluinos, whereas those with a vector
structure are induced by box and penguin diagrams. For each of these
operators there are several similar ones, with different helicity or
colour structure. Those with opposite helicity structure, obtained
through the exchange $L\leftrightarrow R$, are called hereafter
``primed'' operators.

The four-quark operators in~(\ref{penguinboxop}) and~(\ref{boxop}) are
formally of higher order in the strong coupling than the magnetic and
chromomagnetic operators.  The scalar/tensor operators mix at one loop
into the magnetic and chromomagnetic operators, giving rise to a large
log $L$.  Given this fact, the necessity of including 
${\cal O}_{7c,\tilde{g}}$ and ${\cal O}_{8c,\tilde{g}}$ in the operator
basis becomes clear immediately.  Due to these mixing effects, the
scalar/tensor operators have to be included in a LO calculation for the
decay amplitude.  The remaining four-quark operators with vector
structure ${\cal O}_{11,\tilde{g}}^q$--${\cal O}_{14,\tilde{g}}^q$ (and
the corresponding primed operators) do not mix at one loop neither into
the magnetic and chromomagnetic operators nor into the scalar/tensor
four-quark operators.  Therefore, these vector four-quark operators
become relevant only at the NLO precision.
%
\section{Wilson Coefficients at the Decay Scale}
At the electroweak scale, the Wilson coefficients of the various
operators are obtained by perturbation theory; gluino exchanges such as
in fig. 1 contribute. The renormalization program then allows to
calculate the coefficients at the relevant scale $ m_b$.  As already
mentioned, the two terms ${\cal H}_{eff}^{W}$ and 
${\cal H}_{eff}^{\tilde{g}}$ in the effective Hamiltonian~(\ref{hfull})
undergo separate renormalization. The anomalous-dimension matrix of the
SM operators ${\cal O}_1$--${\cal O}_8$ and the evolution of the
corresponding Wilson coefficients to the decay scale $\mu_b$ are very
well known and can be found in~\cite{BG}.  The evolution of the
gluino-induced Wilson coefficients $C_{i,\tilde{g}}$ from the matching
scale $\mu_W$ down to the low-energy scale $\mu_b$ is described by the
renormalization group equation:
\begin{equation}
\label{RGE}
\mu \frac{d}{d\mu} C_{i,\tilde{g}} =  C_{j,\tilde{g}}(\mu) \, 
 \gamma_{ji,\tilde{g}}(\mu) \,.
\end{equation}
The usual perturbative expansion for the initial conditions of 
the Wilson coefficients, 
\begin{equation}
 C_{i,\tilde{g}} (\mu_W) =       C_{i,\tilde{g}}^{0}(\mu_W) + 
 \frac{\alpha_s(\mu_W)}{4\pi} \, C_{i,\tilde{g}}^{1}(\mu_W)  + .....
\label{coeffdecomp}
 \,,
\end{equation}
as well as for the elements of $ \gamma_{ji\,\tilde{g}}(\mu)$, 
\begin{equation} 
 \gamma_{ji,\tilde{g}} (\mu) = 
   \frac{\alpha_s (\mu)}{4 \pi}      \, \gamma_{ji,\tilde{g}}^{0}
 + \frac{\alpha_s^2(\mu)}{(4 \pi)^2} \, \gamma_{ji,\tilde{g}}^{1}
 + .....
 \,,
\label{anomaldecomp}
\end{equation}
is possible because of the choice of including appropriate powers of 
$g_s(\mu)$ into the definition of the operators 
${\cal O}_{i,\tilde{g}}$, as discussed previously.
The anomalous-dimension matrix $ \gamma_{ji,\tilde{g}}$ 
is then a $112\times 112$ matrix. However, the vectorlike character of QCD
and dimensionality arguments reduce it to two identical $54\times 54$ 
matrices. At LO, a further reduction to $8\times 8$ matrices occurs.
At the low scale, the LO expression for the Wilson coefficients of the 
dimension-six operators are
\begin{eqnarray}
{C_{7b,\tilde{g}}(\mu_b)}           & = &  
  \eta^{\frac {39}{23}} 
 \,{C_{7b,\tilde{g}}}(\mu_W) + 
{\displaystyle \frac {8}{3}}  
\left( \eta^{\frac {37}{23}} -\eta^{\frac {39}{23}}\right)
 \,{C_{8b,\tilde{g}}}(\mu_W) + {R_{7b,\tilde{g}}}(\mu_b)\,, 
\nonumber \\ 
{C_{8b,\tilde{g}}(\mu_b)}          & = &  
 \eta^{\frac {37}{23}}
 \,{C_{8b,\tilde{g}}}(\mu_W) + {R_{8b,\tilde{g}}}(\mu_b)\,.
\label{evoldimsixb}
\end{eqnarray} 
where the remainder functions 
$R_{7b,\tilde{g}}(\mu_b)$ and $R_{8b,\tilde{g}}(\mu_b)$ contain
the coefficients of the scalar/tensor operators and are small
in this case. At the NLO, also the operators with vector Lorentz 
structure enter.

\begin{figure}[p]
\begin{center}
\leavevmode
\epsfxsize= 10.5 truecm
\epsfbox[18 167 580 580]{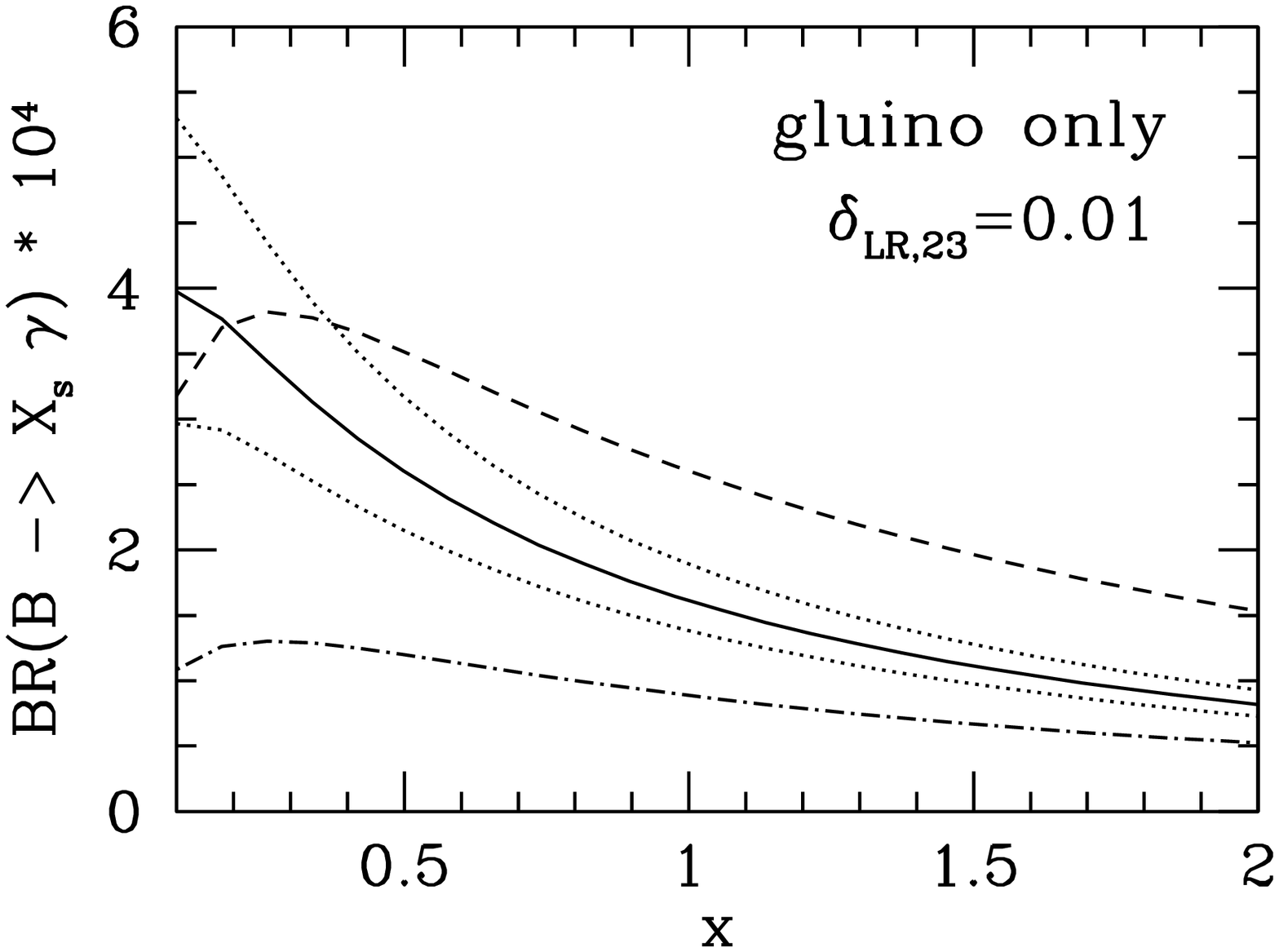}
\end{center}
\vspace*{-0.5truecm}
\caption[f1]{Gluino-induced branching ratio
 ${\rm BR}(\bar{B}\to X_s\gamma)$ as a function of
 $x= m^2_{\tilde{g}}/m^2_{\tilde{q}}$, obtained when the only source
 of flavour violation is $\delta_{LR,23}$ (see text), fixed to the
 value $0.01$, for $m_{\tilde{q}}=500\,$GeV. The solid line shows the
 branching ratio at the LO in QCD, for $\mu_b =4.8\,$GeV and
 $\mu_W = M_W$; the two dotted lines indicate the range of variation
 of the branching ratio when $\mu_b$ spans the interval
 $2.4$--$9.6\,$GeV. Also shown are the values of
 ${\rm BR}(\bar{B}\to X_s\gamma)$ when no QCD corrections are included
 and the explicit factor $\alpha_s(\mu)$ in the gluino-induced
 operators is evaluated at $4.8\,$GeV (dashed line) or at $M_W$
 (dot-dashed line).}
\label{sizeqcd23lr}
%
\begin{center}
\leavevmode
\epsfxsize= 10.5 truecm
\epsfbox[18 167 580 580]{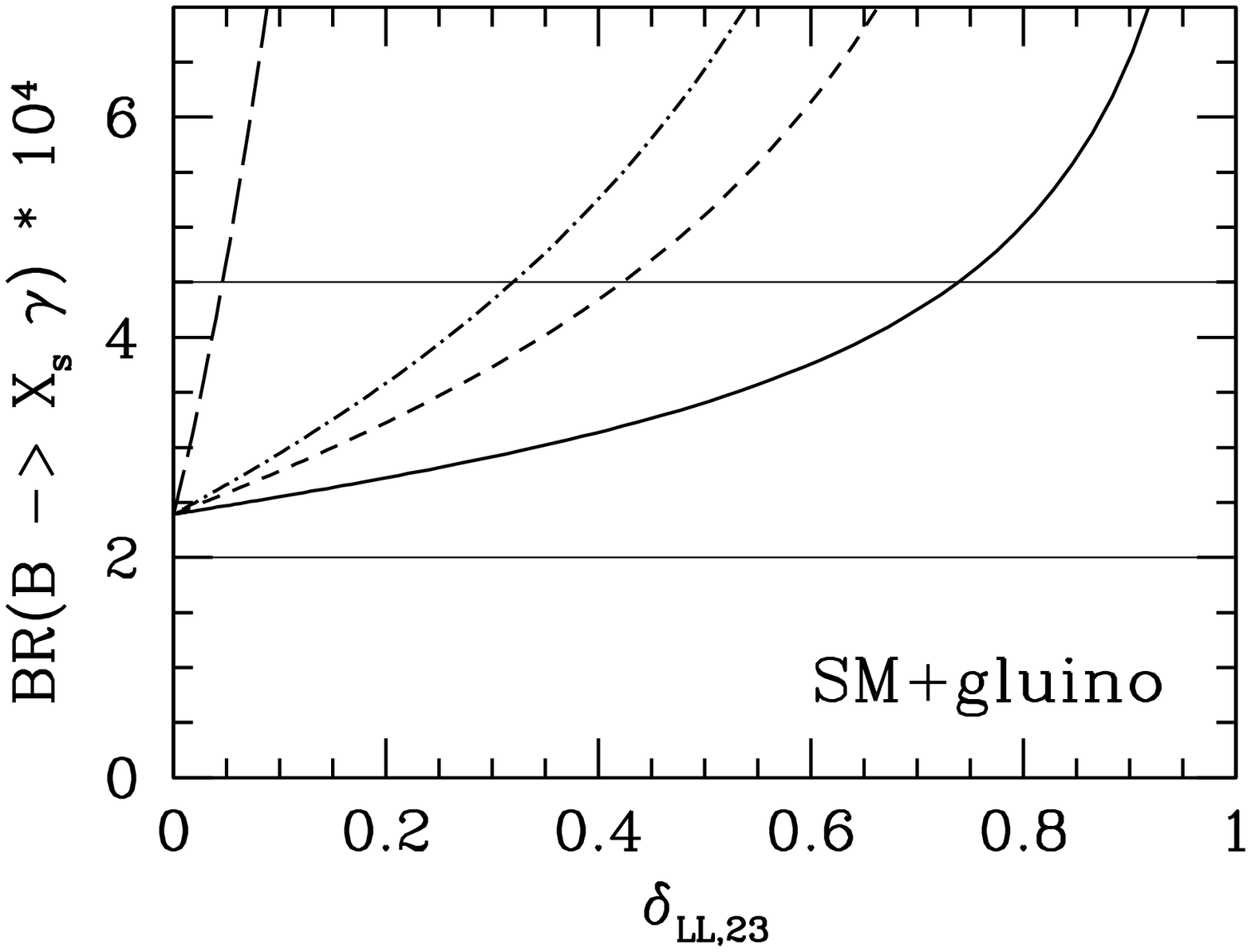}
\end{center}
\vspace*{-0.5truecm}
\caption[f1]{ ${\rm BR}(\bar{B}\to X_s\gamma)$ vs. $\delta_{LL,23}$, 
 when $\delta_{LL,23}$ and
 $\delta_{LR,33}$ are the only sources of chiral-flavour violation.
 The dependence on $\delta_{LL,23}$ is shown for different values of
 $\delta_{LR,33}$: 0 (solid line), 0.006 (short-dashed line), 0.01
 (dot-dashed line), 0.1 (long-dashed line). The value of 
 $x= m^2_{\tilde{g}}/m^2_{\tilde{q}}$ is fixed to $0.3$ and 
 $m_{\tilde{q}}$ to $500\,$GeV. The vertical band indicates the
 experimental constraint.}
\label{glsm23llnew}
\end{figure}
%
\section{Flavour violation and results}
While the main purpose of this talk was to stress the importance of
including and systematically treating the new operators arising in
beyond the SM physics, a few specific results should also be mentioned.
Gluino exchange gives rise to flavour changes if the quark and squark
mixing matrices are different (more precisely: do not commute).  This
happens when the so-called soft contributions to the squark mass matrix
are substantially off-diagonal. Of course the values depend on the
specific model; here we introduce phenomenologically the quantities~(see
ref.~\cite{GGAS}):
\begin{equation} 
\delta_{LL,ij} = \frac{(m^2_{\,d,\,LL})_{ij}}{m^2_{\tilde{q}}}\,, 
\hspace{1.0truecm}
\delta_{RR,ij} = \frac{(m^2_{\,d,\,RR})_{ij}}{m^2_{\tilde{q}}}\,, 
\hspace{1.0truecm}
(i \ne j) 
\label{deltadefa}
\end{equation}
which are a measure for flavour violation if $i \neq j$. The 
various $m^2$ are the elements of the soft mass matrices. The parameters
are varied, as usual, in their allowed ranges.

Figs. 2 and 3 exemplify the results of this work. While the first
shows how much systematic QCD corrections influence the branching ratio,
fig. 3 illustrates its sensitivity to the soft matrix elements.
%
\ack
We thank our collaborators Ch. Greub and T. Hurth for a fruitful
collaboration. This work was partially supported by the
Schweizerischer Nationalfond.

\end{document}